# Efficient Android Based Invisible Broken Wire Detector

V. Jegathesan[1], T. Jemima Jebaseeli[2], D. Jasmine David[3]

[1]Associate Professor, Department of Electrical and Electronics Engineering, Karunya Institute of Technology and Sciences, Karunya Nagar, Coimbatore, Tamil Nadu, India. Phone: 8778256881, Email: jegathesan@karunya.edu
[2]Assistant Professor, Department of Computer Science and Engineering, Karunya Institute of Technology and Sciences, Karunya Nagar, Coimbatore, Tamil Nadu, India.Phone: 9487896556, Email: jemima_jeba@karunya.edu
[3]Assistant Professor, Department of Electronics and Communication Engineering, Karunya Institute of Technology and Sciences, Karunya Nagar, Coimbatore, Tamil Nadu, India. Phone: 8838083460, Email: jasmine@karunya.edu

**ABSTRACT**

The system to include an android based underground broken wire detection which works by detecting the electromagnetic field around a live underground cable is proposed. The transmission cables undergo stress and strain as they are under the ground. This may lead to short circuits or various kinds of snapping in the wire. If these faults are not treated, it may cause an interruption in the power supply and permanent damage. The proposed method distinguishes the short circuit fault in the underground links. The existing and traditional techniques for detection are reviewed and only the methods for spotting the short circuit error are included. Thus the proposed system provides a cost-efficient way of detecting the short circuit shortcomings in underground cables.

## 1. INTRODUCTION

In today's world electricity has become a major part. Considering the development in the past years, it is known that this world has evolved a lot as new technologies came up and the old and basic ones were upgraded [2, 15]. The discovery of electricity has been of greater importance as it provided power supply and also acted as a source of recharge for many gadgets [13]. With all these, the power had to be supplied from one place to the other [4, 5, 9] with low power losses and the ability to absorb emergency power loads [12]. Overhead transmission evolved very fast but due to its disadvantages, the underground transmission is used to a larger extent [1].

The electrical power system comprises of producing framework, transmission framework, and conveyance framework in which the transmission framework goes about as a connection between the generating framework and appropriation framework [3]. Transmission of current should be possible by two sorts of transmission lines, overhead transmission lines which are visible as they are placed on poles above ground, cheap, easy to install, troubleshoot, and upgrade, therefore it is widely used in the power systems [7]. Despite all of this it cannot be installed in densely populated zones and missing link areas. To beat this, underground transmission lines are utilized, they have lower permeability and less influenced by climate; subsequently this is the most well-known part in urban regions [6, 14]. Anyway the lower permeability of the underground transmission line makes trouble keep up. The way of the cable wire must be known if the beginning stage is realized. But it is difficult to analyze the entire link along the length if there is any shortcoming happening in it [8, 10, 11].

In this paper a kind of transmitter and recipient technique is presented in which the identifier circuit is utilized to distinguish the short out issue in the link. The principle motivation behind this paper is to construct a circuit that goes about as a gadget to identify the situation of a messed up purpose of the conductor inside the link along these lines limit the wastage of assets just as spares time. The word invisible interprets the wire that is not visible to the eyes or it is far from the touch which directly indicates to underground wires or the wires in the walls. The basic point that is taken into consideration is to detect the exact position of the breakage in the cable. The device is used to detect the electromagnetic field around the cable and the frequency in the cable and indicate if there is any interruption in the flow of current in the cable. The device moving on the surface of the earth by following the flow of current in the cable and indicate if there is any interruption or breakage.

## 2. METHODOLOGY

The proposed paper incorporates both hardware and software parts. The fundamental reproduction work is done in simple EDA programming. The product part likewise incorporates Arduino which is one of the fundamental controller circuits.

The equipment part overwhelmingly contains the identifier unit and the motion observing unit. This venture on an entire depends on two phases namely, programming execution and equipment development.

As shown in Figure 1, the main process of the detector device is to detect the conductor fault in the cable such as breakage of a conductor by short circuit fault and external mechanical faults. There is an antenna used in receiving the electromagnetic radiation and an oscillatory circuit that acts as a transmitter. The frequency range of the oscillatory should match the range of the frequency of the cable that is received by the probe which is a part of the device. The gesture monitoring unit contains a microcontroller that is used to connect all the parts and form a single unit. On the other hand an android device is used to control the movement and direction of the device.

**2. 1 Algorithm of gesture monitoring unit**

1. Start
2. Create a robot control program using Arduino for forward, backward, left, and right drive of the robot.
3. Upload the program to the Arduino.
4. Turn on Bluetooth in the android device.
5. Connect with the accessible Bluetooth module and pair them using the HC-05 BT module.
6. If not available turn off and on the Bluetooth again to pair with the gadget.
7. Open the app in the android device using an android Bluetooth controller app.
8. Select the HC-05 BT module and click on controller mode.
9. Set the values to the arrows for movement according to the program and save them.
10. Move the robot according to the detector module direction.
11. Stop

**2.2 Algorithm of detector module**

1. Start
2. Initializing the movement of the robot
3. Letting the detector module pick up high voltage AC using the probe of magnetic flux link conductor to induce a voltage in it.
4. Sense the high voltage AC that allows the oscillator circuit to oscillate.
5. Oscillator circuit output to determine the transistor to conduct which glows the LED.
6. Probe avoids the high voltage AC then the diode directs and hinders the oscillator Circuit from wavering not letting LED sparkle.
7. Stop

**2.3 Cable fault detector unit**

The cable fault detector consists of two parts. They are, the detector unit and gesture monitoring unit. The main objective of the proposed work is to detect the breakage of wire by designing a detector unit which moves on the surface of the earth with the help of a gesture monitoring unit which places the major role in the movement of the device.

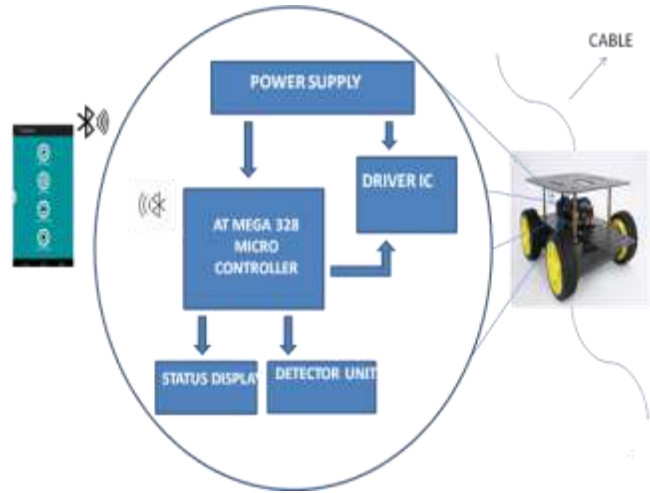

**Figure 1.** Block Diagram of the proposed system.

The identifier unit assumes a significant job in distinguishing the shortcoming in the cable. In the underground cables, there will be current flows through the cable. Due to this, there will be an electromagnetic field around the cable. As there is an electromagnetic field in the cable there will be a frequency of range 50-60Hz which is the common range. The range of electromagnetic radiation depends on the depth of burying the cables. The more depth the cable is buried the less the range of radiation. That means if the cable is near the surface the radiation is more. While look at the detector unit, it has an oscillatory circuit that is used to find the frequency range which is coming from the underground cable. Hence, there will be an electromagnetic radiation around the cable where ever it is laid in the ground. This implies that the current flow through the cable has to be identified by the oscillatory circuit by matching the frequency.

The circuit diagram for the detector unit is given in Figure 2. The main segment of the detector unit is IC-CD4069 which has a voltage range of 3.0V to 15V, low power consumption, and high noise immunity. It is a hex inverter CMOS IC comprises of six inverter circuits. It will help in detecting the electromagnetic field. The diverse number and values of diodes, resistors, capacitors, and transistors aid in improving the range of detection and switching purposes. There is an oscillatory circuit in the detector unit and the equations given below are used to adjust the frequency range and match with the range of the cable.

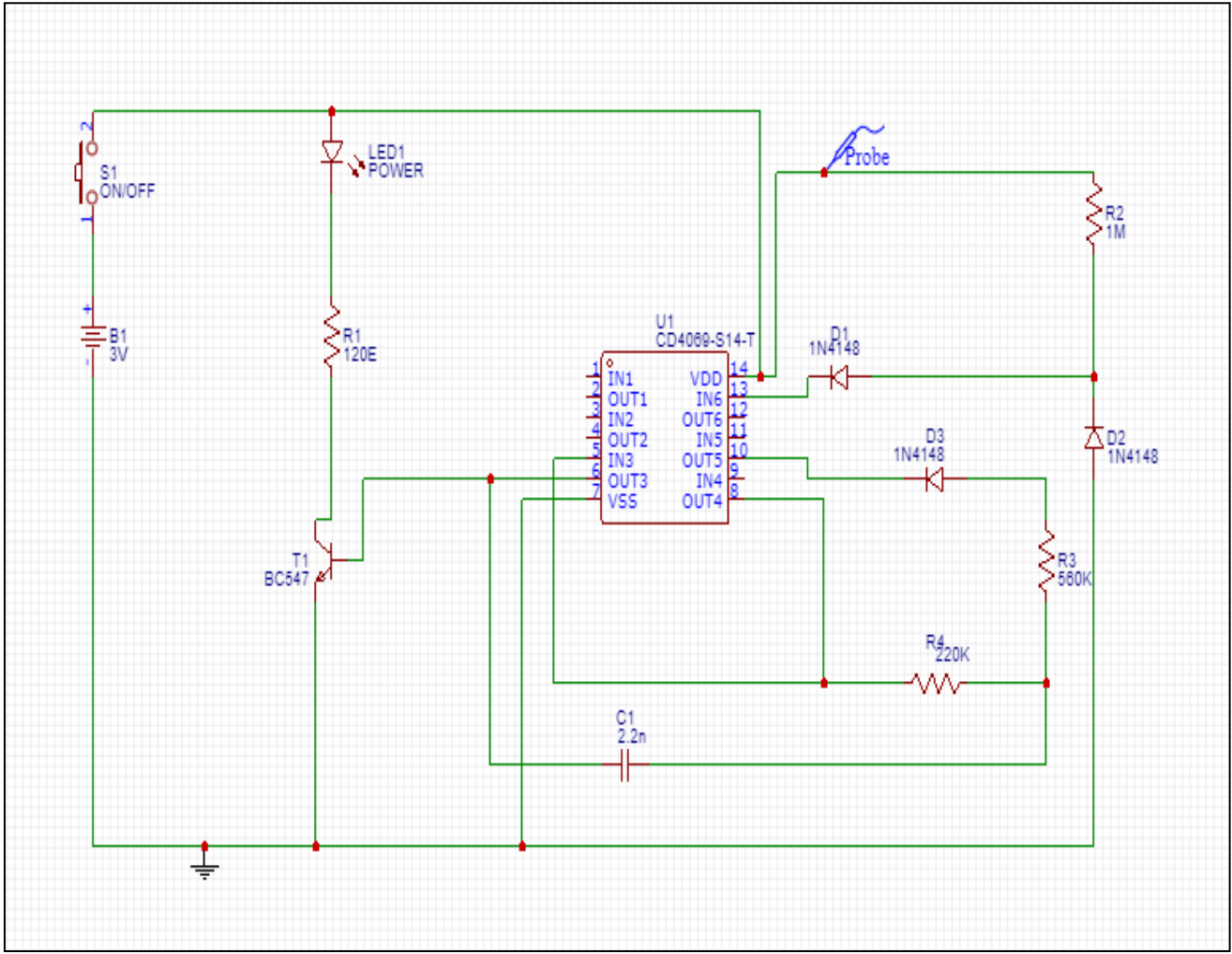

**Figure 2.** The circuit diagram of the detector unit.

$$T = Rc \ln \frac{(V_{DD}+V_D)^2}{V_T(V_{DD}-V_T)} + RC \frac{K}{1+K}\left[\ln \frac{K(V_{DD}+V_T)+(V_T-V_D)}{K(V_{DD}+V_D)} + \ln \frac{K(2V_{DD}-V_T)+(V_{DD}-V_T-V_D)}{K(V_{DD}+V_D)}\right]$$

(1)

$$K = \frac{R_S}{R} \quad (2)$$

$$F = \frac{1}{T} \quad (3)$$

Where Rs is the series resistor
R – Timing resistor
C – Timing capacitor
$V_{DD}$ – Power supply voltage
$V_D$ – IC internal protection diode forward voltage
$V_T$ – Inverter threshold voltage

The equations provide a useful prediction of oscillator frequency when the values of $R$ (Timing Resistor), $R_S$ (Series Resistor), and $C$ (Timing Capacitor) are inside sensible limits.

For the most part, the conditions give the most exact outcomes when;
1. The oscillator periods are generally enormous contrasted with the engendering and change defers innate inside the coordinated circuit.
2. C (C1) is moderately enormous contrasted with the inherent capacitances inside the coordinated circuit in the physical design.

3. R (R3) is sufficiently enormous to permit the inverter's yield to swing near the force flexibly rails.

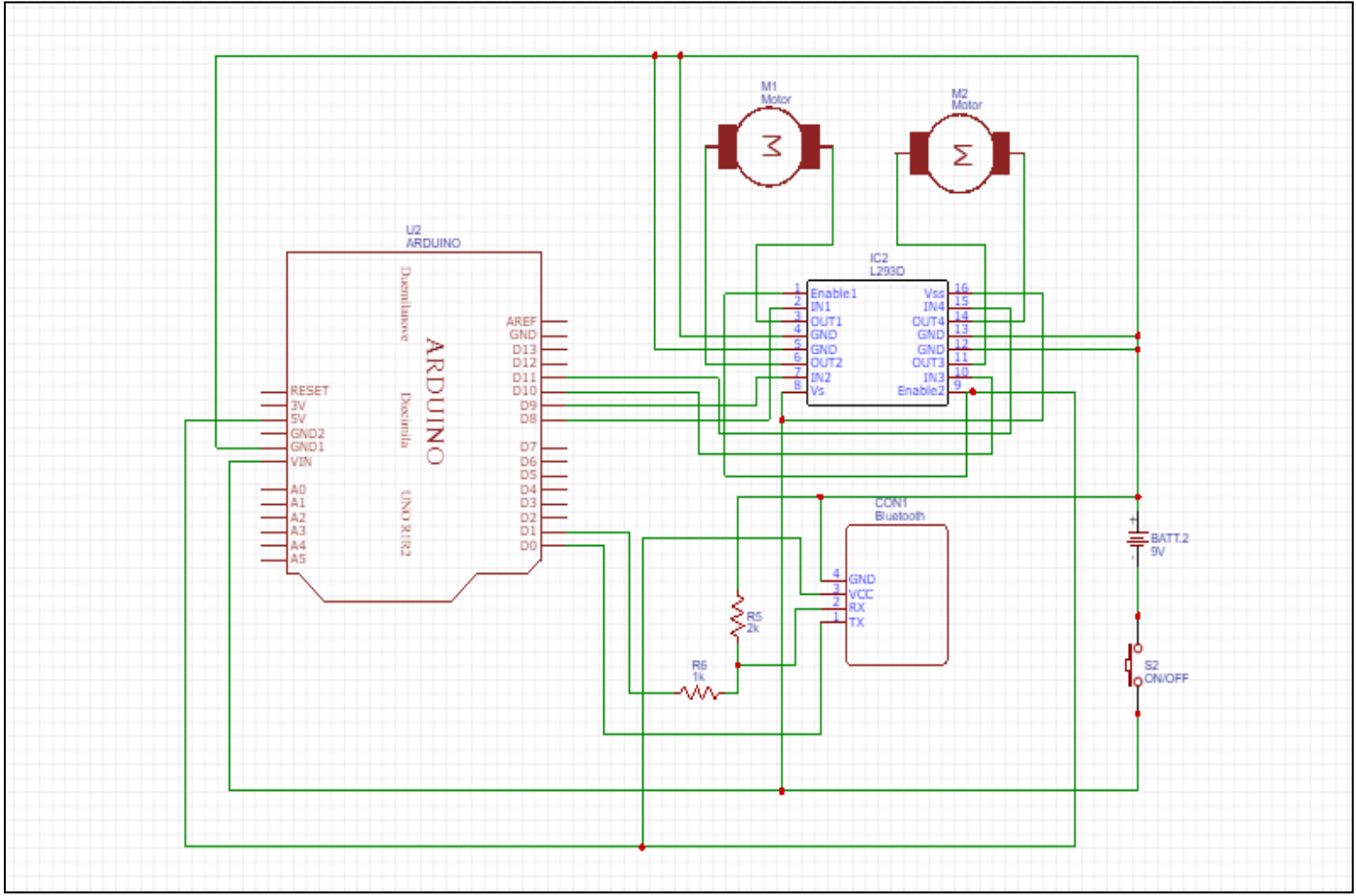

**Figure 3.** The circuit diagram for the gesture monitoring unit.

4. $R_S$ (R4) is sufficiently little to forestall a phase shift inside the oscillation criticism way because of the input capacitance of the inverter, yet huge enough to forestall the inverter's information clasping diode from stacking the input organize. The suggested values are in the range of two and multiple times R.

**2.4 Gesture monitoring unit**

The Figure 3 shows the circuit diagram for the gesture monitoring unit along with the motor driver. The gesture monitoring unit constitutes of the microcontroller, driver circuit, and Bluetooth interfaced together. The microcontroller employed here is Arduino Uno -ATmega328P. It is the main controller of this unit and avails in controlling the driver circuit, motors, and Bluetooth by the coding uploaded to it. The driver circuit uses the IC L293D which operates on the H-Bridge concept moves the two motors simultaneously in any direction that helps to move the robot without much delay. The motors connected to the driver circuit rotate the wheels and enable the robot to move. The Bluetooth module connected to Arduino for wireless communication with the android device is JY MCU HC-05. The app installed in the android device for connection with the Bluetooth module is Arduino. The direction control is set by the Bluetooth controller. This app helps in connecting the hardware with the phone and also controlling it. The main benefit of using this Bluetooth module is that it works both in master and slave configurations whereas the other modules have either one of master or slave operation.

**2.5 Types of faults**

There are different sorts of issues happening in underground links that ought to be known before finding out about the techniques for flaw recognition. The continuous issues happening in underground links are as follows,

- Open circuit issue
- Short circuit issue
- Earth issues

Most of the errors occur when moisture enters the insulation, mechanical trauma during transferring from one place to another, laying process, and multiple strains experienced by the cable through its working life.

## 2.6 Underground cable layout

The underground cables are usually laid at a depth of 25 inches. There are different methods in which the cables are laid under the ground. Based on the safety issues the burying of cables is done in a very organized manner through different methods are as follows,
- Direct laying
- Draw-in system
- Solid system

The most common and easy method used is the direct laying method in which the cables are placed inside the trenches. These trenches are concrete pits in which the cables are placed in and filled with sand for safety purposes. These trenches are covered with a concrete slab.

## 2.7 Path of the cables

The trenches are made on the sides of the road without causing any disturbance. In most of the areas, these tranches are made visible so that there is no need to dig the road. And these trenches are marked by signs. The detector device is placed on the surface direction to find the path of underground cables. The following are the techniques used for fault detection.

### 2.7.1 Online method

Online technique uses the inspected current to manage the deficiency focuses.

### 2.7.2 Disconnected method

This technique utilizes extraordinary instruments to try out the assistance of links in the field. This strategy is additionally characterized into two strategies. Example: tracer technique and terminal technique.

### 2.7.3 Tracer method

This strategy works together in discovering the issue of the link by strolling on the link lines. The flaw area is represented from an electromagnetic sign or audible sign. This strategy is utilized to discover the issue area precisely.

### 2.7.4 Terminal method

The terminal technique is used to analyze the situation of the shortcoming in a link from one end or both the closures without following. This strategy is utilized to discover general territories of the shortcoming to invigorate following on underground link.

There are a few other traditional methods which are used for fault detection in submerged cables. Thumping is a process in which the high voltage is supplied to a faulted cable, resulting in the production of high-current arc that makes a noise loud enough to be heard above the ground making it possible for fault detection. Sectionalizing is another process that involves physically cutting and splicing the cable to smaller pieces and finding the fault. Lastly the Time Domain Reflectometry (TDR) procedure changes the cable impedance when a fault is produced; thereby affecting the ability to transmit pulses it also enables a tester to calculate the distance to changes in a cable.

## 3. HARDWARE IMPLEMENTATION

As shown in Figure 4, there is a probe which is connected to the IC-CD4069. This probe is made out of copper metal and acts as an antenna in receiving the signals from the cable. These signals are electromagnetic radiations emitted to the surroundings from the cable.

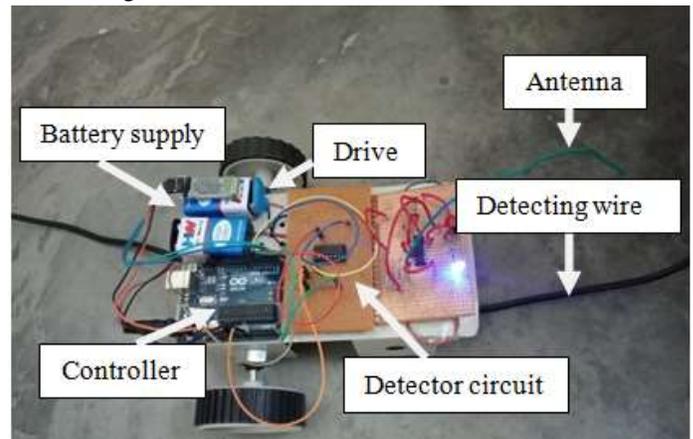

**Figure 4.** The hardware implementation of the system.

According to the requirement if the frequency of the emitting cable is low the length and thickness of the probe are to be adjusted and made higher. Underground cable fault detection makes it easy for tracing the exact fault location and distance. It is more compact and reliable thus helps in saving time. As mentioned earlier, it is cost-efficient. Therefore it helps in reducing unnecessary expenses. Also, it requires low maintenance and operating cost as the damage rate is low.

For experimental purposes, the cable kept below the detector unit and 2-meter cable was used. The cable is made as open after 1.5m. The cable was connected with the power line. The proposed experimental research is made to pass through the testing cable. Readings are taken for every 0.5m distance. Since the open cable is made after 1.5m length of the cable, the frequency of the electromagnetic field is obtained and it is indicated as No-fault condition. For the experimental setup, the distance between the cable and detector is less, and the frequency obtained is almost near to 45Hz. As shown in Table 1, if the depth of the cable is more, then the frequency of the electromagnetic field would be of lesser value. The open fault is executed after 1.5m length of cable. After 1.5m length of cable, there is no electromagnetic field obtained and hence the fault condition was considered as Yes.

**Table 1** Experimental result with fault conditions.

| S. No | Distance in meter | Frequency in Hz | Fault condition |
|---|---|---|---|
| 1 | 0.5 | 44.5 | No |
| 2 | 1.0 | 45.1 | No |
| 3 | 1.5 | 44.3 | No |
| 4 | 2.0 | 0 | Yes |

The obtained results are implemented using the following graph and it is shown in Figure 5. It enables the system to find the status of the process until the fault is identified; the frequency of the electromagnetic field is almost constant. Once the open fault is identified, the frequency of the electromagnetic field is dropped to a very low value.

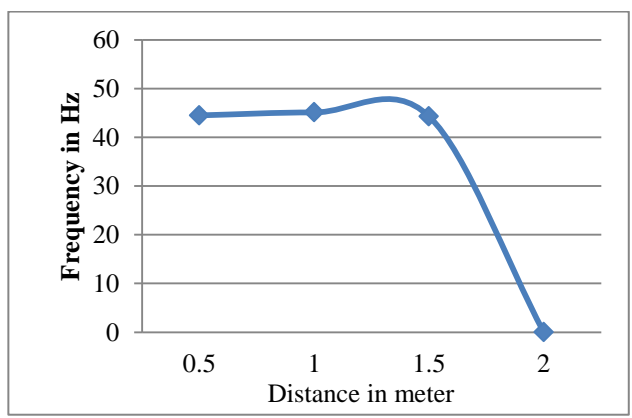

**Figure 5.** The statistical result of electromagnetic field when open fault is identified.

The proposed detector unit is suitable for congested urban areas as it reduces live-wire contact injuries, improves public safety, produces no danger to wildlife or low flying aircraft, ensures small voltage drops, not easy to steal and vandalize, and avoids the chances of illegal connections. As the cables are under the ground, fault interruptions are very less and these cables are not responsive to shaking and shorting due to vibrations, wind, accidents, etc. This makes it necessary to detect the fault in underground cables.

## 4. CONCLUSION

The proposed work centers on the purpose of recognizing the shortcoming in any AC carrying cable which is buried under the ground. If there is an occurrence of short circuit fault in any AC line ranging up to 440V, then it is detected with the help of the detector unit. The uniqueness of the system lies in the detector circuit as it mainly focuses on the frequency range and doesn't involve any sensors. This makes the equipment cost proficient and the utilization of android gadgets makes it easier to use and advantageous to work.